\documentclass[11pt]{article}

\usepackage{graphicx}

\begin{document}

Version Date: May 30th, 2005




\bigskip

Title:\\
{\textbf{Fast, Preisach-like  characterization  of  hysteretic 
 systems}}

\bigskip

Authors:\\
Marek W.~Gutowski$^{1}$, Lajos K.~Varga$^{2}$ and Attila
K\'akay$^{2}$

\bigskip

${}^{1}$ Institute of Physics,  Polish  Academy 
of Sciences, Al.~Lotnik\'ow 32/46, 02--668 Warszawa, Poland\\
${}^{2}$ Research Institute for  Solid  State  Physics  and 
Optics,  Hungarian Academy of Sciences, P.O.Box 49, H-1525 Budapest, Hungary

\bigskip

email: Marek.Gutowski@ifpan.edu.pl



\bigskip

\begin{abstract}
    Proposed is a  substantially  simplified,  Preisach-like  model  for 
 characterization  of  hysteretic  systems,   in   particular   magnetic 
 systems. The  main  idea  is  to  replace  a  two-dimensional  Preisach 
 density with just two real functions, describing in  a  {\em  unique\/} 
 way the reversible and irreversible processes. As a  byproduct  of  our 
 model we prove, that the major hysteresis loop  alone  is  insufficient 
 to produce the unique Preisach map.
\end{abstract}

\bigskip



\newpage

\section{Introduction}

    One of the most widely known classes of  hysteresis  models  is  the 
 Classical Preisach  Model  (CPM)  and  a  rich  variety  of  its  close 
 relatives.

    The  main  motivation  behind  our  work  was  clear  separation  of 
 reversible and irreversible  components  of  isothermal,  quasi  static 
 hysteresis loops.  In  CPM,  the  reversible  part  is  represented  by 
 hysterons located on the edge of the Preisach half-plane, i.e.  on  the 
 line  $H_{\uparrow}=H_{\downarrow}$  (up-  and  down-switching   fields 
 equal to each other). All other physically valid hysterons, i.e.  those 
 with $H_{\uparrow}>H_{\downarrow}$, are  related  to  the  irreversible 
 magnetization  processes.  Usually,  the   Preisach   distribution   is 
 interpreted in probabilistic terms. There is, however, a  problem  with 
 credible  characterization  of   the   (ferro)magnetic   samples   with 
 diamagnetic  components,  like,   for   example   the   graphite   foam 
 \cite{graphite}, and -- of course --  superconductors.  Their  Preisach 
 maps {\em  may\/}  go  negative  in  the  reversible  part,  thus  such 
 processes should be treated separately from the rest  of  the  complete 
 Preisach distribution. Disregarding this distinction  leads  frequently 
 to puzzling results: the negative amplitudes  appear  in  the  Preisach 
 plane, even for ferromagnetic samples. Findings  like  that  have  been 
 signalled since at least ten years. Until recently, they  were  usually 
 attributed to
\begin{figure}[f]
\centerline{\includegraphics[scale=0.33,angle=-90]{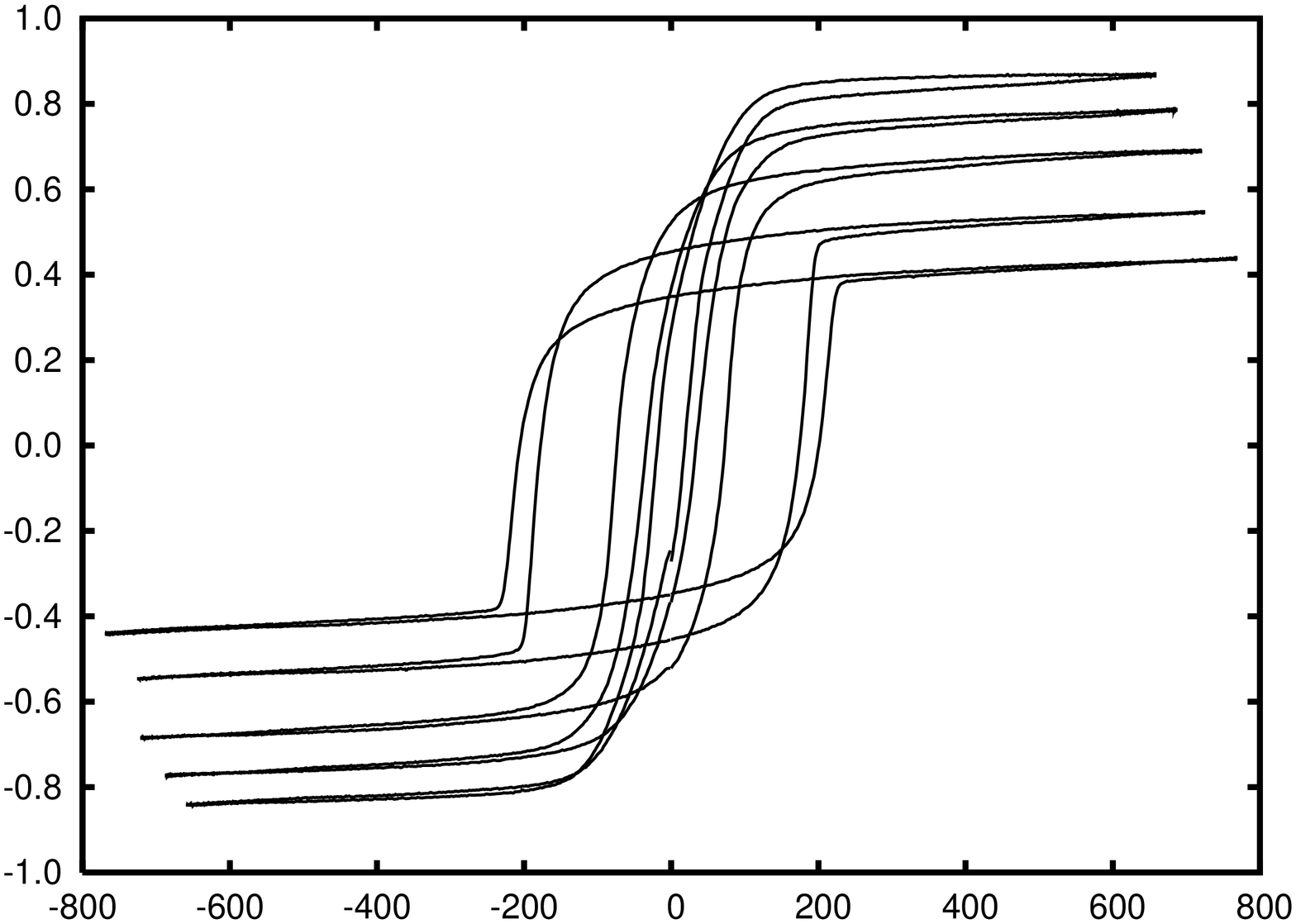}}
    \caption{Room temperature major hysteresis loops  of  FeZrB$_{12}$Cu 
 nanocrystalline material annealed at:  $100^{\circ}$C,  $180^{\circ}$C, 
 $240^{\circ}$C,   $360^{\circ}$C   and    $440^{\circ}$C.    Coercivity 
 increases with increasing annealing temperature.}
\label{fig1}
\end{figure}
    experimental uncertainties or  inaccuracies  introduced  during  the 
 data processing.  Yet,  even  the  FORC  diagrams  introduced  by  Pike 
 \cite{Pike}, which are more or less equivalent  to  CPM,  exhibit  such 
 a~feature quite  often.  Moreover,  the  very  existence  of  composite 
 materials  with  negative  remanence  \cite{negative}  constitute   the 
 unquestionable   experimental   evidence,   that   the    probabilistic 
 interpretation of the Preisach distribution is untenable.

\section{Model description}

    Inspecting results, similar to those presented by Pike and by  other 
 authors \cite{Stancu},  we  decided  to  try  the  following  idea.  We 
 drastically  reduce  the  support   of   the   Preisach   distribution, 
 $\varrho(H_{\uparrow},H_{\downarrow})$,\ leaving for it only two lines: 
 $H_{\uparrow}=H_{\downarrow}$     for     reversible     part,      and 
 $H_{\uparrow}=-H_{\downarrow}$, the mirror symmetry axis of  the  usual 
 Preisach distribution, for irreversible part. This way we can  separate 
 both kinds of magnetization processes. Formally:

\begin{eqnarray}
    \varrho   \left(   H_{\uparrow},   H_{\downarrow}   \right)   &   =& 
 \varrho_{\rm rev}(H)\cdot \delta  \left(H_{\uparrow}  -  H_{\downarrow} 
 \right) \nonumber \\
    &+& \varrho_{\rm irr}(H)\cdot \delta \left( H_{\uparrow} +
 H_{\downarrow}\right),
\end{eqnarray}
    where  $\varrho_{\rm  rev}(\cdot)$  and  $\varrho_{irr}(\cdot)$  are 
 real,   single-valued   functions   of   a   single    variable,    and 
 $\delta(\cdot)$ is Dirac's delta.

\begin{figure}[f]
\centerline{\includegraphics[scale=0.33,angle=-90]{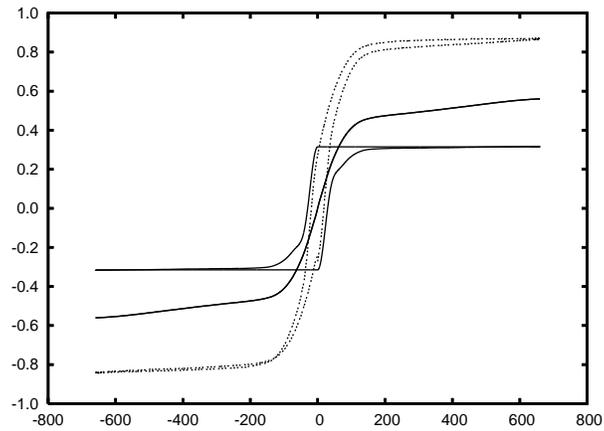}}
    \caption{Decomposition of the major hysteresis  loop  (dotted  line) 
 into reversible  and  irreversible  components  (solid  lines).  Sample 
 annealed at $100^{\circ}$C.
 }
\label{fig2}
\end{figure}
\begin{figure}[f]
\centerline{\includegraphics[scale=0.33,angle=-90]{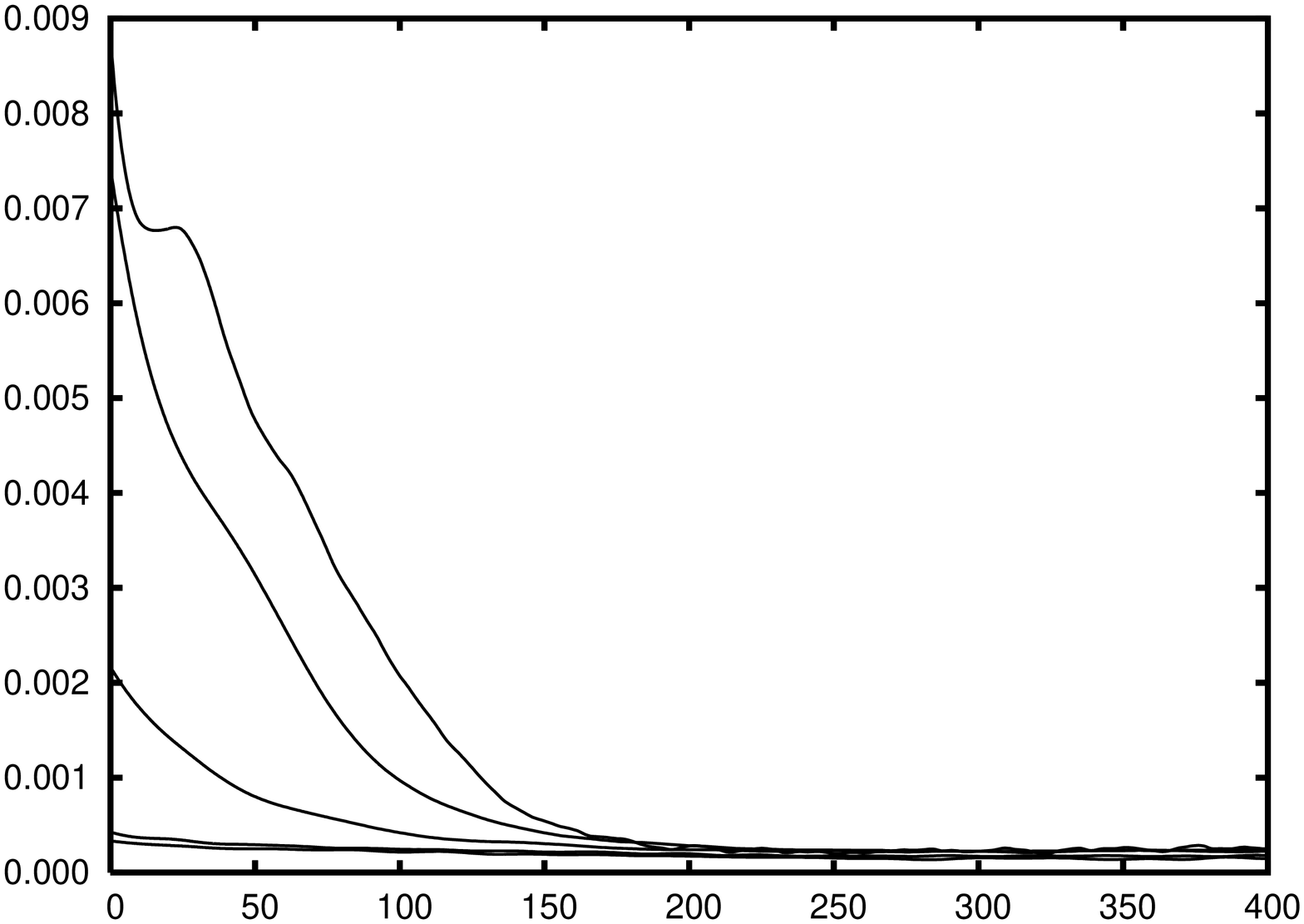}}
    \caption{Reversible parts,  $\varrho_{\rm  rev}(H)$,  of  hysteresis 
 loops shown  in  Fig.~\ref{fig1}  diminish  with  increasing  annealing 
 temperature.   Note   the   composite,   non-Langevin    shape    after 
 low-temperature annealing.}
\label{fig3}
\end{figure}
    In this respect our idea is similar to the one presented earlier  in 
 \cite{Langwagen}. The separate treatment for the reversible  processes, 
 using different arguments than ours, was pointed out  in  many  papers, 
 maybe most convincingly in \cite{Perevertov}.

\subsection{Extracting   $\varrho_{\rm   rev}(\cdot)$   and    $\varrho_{\rm
 irr}(\cdot)$ from experimental data}

    Our  calculations  strictly  follow  the  usual  path  of  all   the 
 Preisach-like models. For $H<0$, the  function  $\varrho_{\rm  rev}(H)$ 
 is immediately identified as $\varrho_{\rm rev}(H)=dM(H)/dH$, with  the 
 derivative  being  taken  along  the  ascending  branch  of  the  major 
 hysteresis  loop,  while  for  $H>0$  we  have  to  use  the   identity 
 $\varrho_{\rm rev}(H)=\varrho_{\rm rev}(-H)$.

    The other function is again given by the  derivative:\\
  $\varrho_{\rm 
 irr}(H\ge 0 )  =  -d/dH  \left[  M_{d}(H)  -  M_{a}(H)  \right]$.  Here 
 $M_{d}$ describes the descending branch of the major  hysteresis  loop, 
 and $M_{a}$ -- its ascending branch.

\subsection{Repairing the unphysical results}

\begin{figure}[f]
\centerline{\includegraphics[scale=0.33,angle=-90]{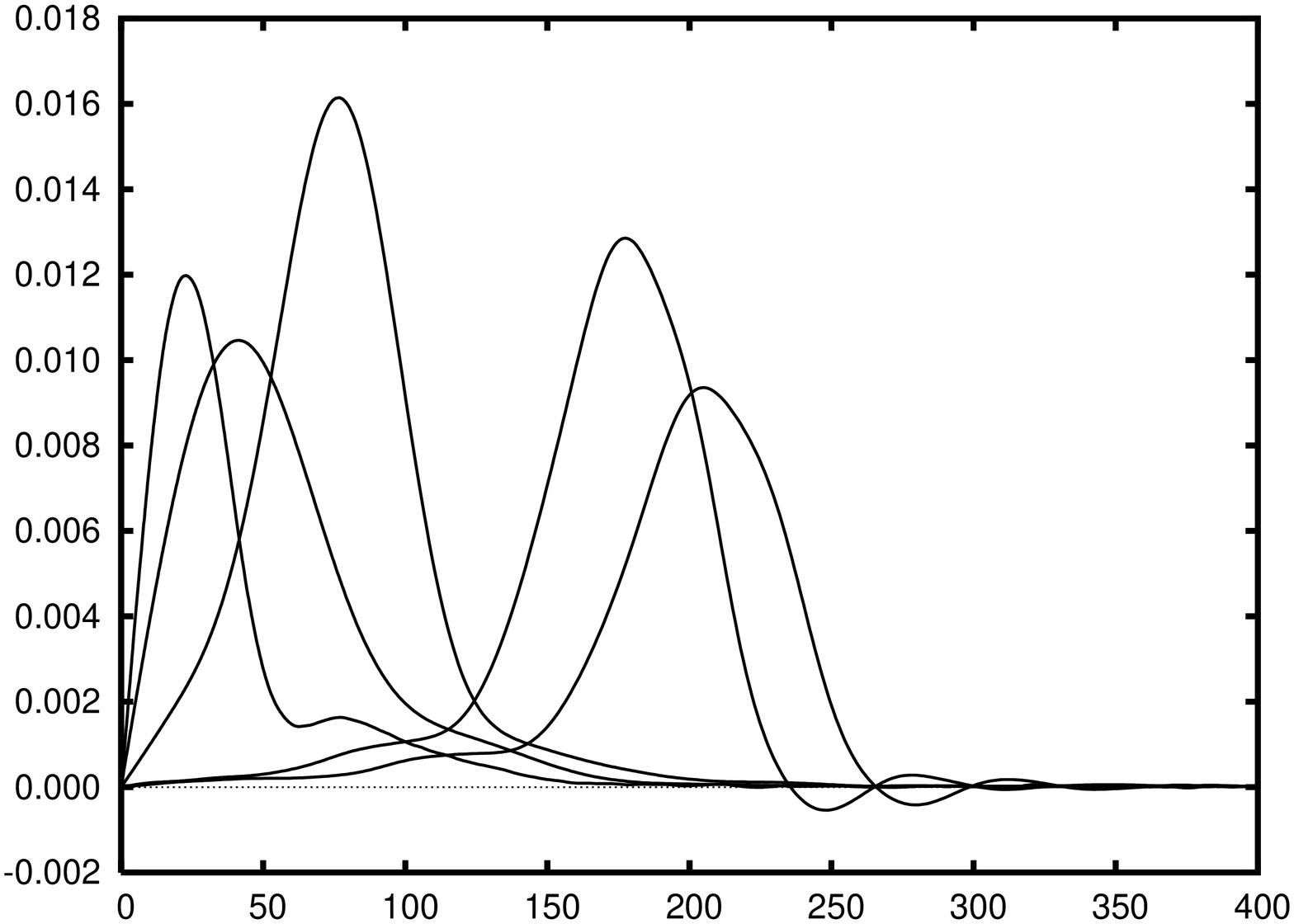}}
    \caption{Irreversible  parts  of  hysteresis  loops,   $\varrho_{\rm 
 irr}(H)$,  shown   in   Fig.~\ref{fig1}.   The   increasing   annealing 
 temperature activates the hysterons with higher  coercivity.  Note  the 
 regions  with  negative  amplitudes  followed  by   ''regular''   ones. 
 Amplitudes  expressed   in   the   same,   arbitrary   units,   as   in 
 Fig.~\ref{fig3}.}
\label{fig4}
\end{figure}

    The experience shows, that sometimes $\varrho_{\rm irr}(H)$  happens 
 to be negative on some interval, see Fig.~\ref{fig4}, near  $H=250$~Oe. 
 This is, as a rule,  accompanied  by  the  presence  of  the  similarly 
 shaped, but positive, part of the function  $\varrho_{\rm  irr}(\cdot)$ 
 located at higher fields. To get  rid  of  the  `unphysical'  parts  of 
 $\varrho_{\rm irr}(\cdot)$, we can apply the relation:
\begin{eqnarray}
Hy(a,-a) - Hy(b,-b) =\nonumber\\
Hy(b,-a) +Hy(a, -b) -Hy(b,b) -Hy(-b,-b)&&
\label{transform}
\end{eqnarray}
    where  $Hy(a,b)$  represents  an  ordinary  Preisach  hysteron  with 
 $H_{\uparrow}=a$ and $H_{\downarrow}=b$. From  the  above  identity  we 
 can conclude, that essentially it might be  possible  to  'repair'  the 
 unphysical results, at the cost of restoration  of  the  full  Preisach 
 model.
    Unfortunately,  this  can  be  done   in   infinitely   many   ways. 
 Accidentally, we have just shown, that {\bf the Preisach  map  obtained 
 from major hysteresis loop alone is not unique\/}.

\section{Advantages and drawbacks of the model}

    First of all, our model is unique -- by construction.  Secondly,  it 
 compares very favorably to  others,  when  it  comes  to  computational 
 complexity,  accuracy  and  experimental   effort   needed   for   data 
 acquisition. We only use first derivatives of the original  data,  what 
 introduces definitely less  noise  than  other  techniques  relying  on 
 higher derivatives. The calculations are  straightforward  and  do  not 
 include any  iterative  schemes.  Given  perfect  data,  the  procedure 
 reproduces them exactly. There is no need to guess, or to  assume,  any 
 particular shape of the anhysteretic part of the magnetization  process 
 first, as it is required in Jiles-Atherton model \cite{JA}.

    The  clearly  visible  drawback  of   our   method   is   that   the 
 (over?)-simplified  Preisach  distribution  cannot  be  interpreted  in 
 usual terms, namely as a~probability  distribution.  As  of  today,  we 
 cannot offer its precise alternative interpretation, in  well  defined, 
 physical terms. The ability of our model to reproduce minor  loops  has 
 not been tested yet.

\section{Discussion}

    The simplified procedure, as being rather simple, fast  and  unique, 
 may be regarded as a new, valuable tool of analysis  of  magnetic  (and 
 maybe  superconducting  as  well)  systems,  including  non-homogeneous 
 ones,  like  those  containing  nanocrystals,  multilayer   structures, 
 patterned media, etc. The results produced with  this  approach  should 
 probably be interpreted in the same spirit, as  those  obtainable  from 
 FORC diagrams, that is, in  the  language  of  the  statistics  of  the 
 effective internal fields, including their signs, not just amplitudes.

    Especially interesting is the shape of $\varrho_{\rm rev}(H)$, sometimes
 exhibiting more than one-peak structure, thus  clearly  deviating  from 
 the  usually  assumed  Langevin's  formula.  We   suspect   its   close 
 relationships   with   internal   field   distributions,   investigated 
 numerically in \cite{Morentin}, for systems  with  long-range,  dipolar 
 interactions.



\newpage
List of figure captions:\\

Fig. 1:\\
 Room temperature major hysteresis loops  of  FeZrB$_{12}$Cu 
 nanocrystalline material annealed at:  $100^{\circ}$C,  $180^{\circ}$C, 
 $240^{\circ}$C,   $360^{\circ}$C   and    $440^{\circ}$C.    Coercivity 
 increases with increasing annealing temperature.

Fig. 2:\\
Decomposition of the major hysteresis  loop  (dotted  line) 
 into reversible  and  irreversible  components  (solid  lines).  Sample 
 annealed at $100^{\circ}$C.

Fig. 3:\\
Reversible parts,  $\varrho_{\rm  rev}(H)$,  of  hysteresis 
 loops shown  in  Fig.~\ref{fig1}  diminish  with  increasing  annealing 
 temperature.   Note   the   composite,   non-Langevin    shape    after 
 low-temperature annealing.

Fig. 4:\\
Irreversible  parts  of  hysteresis  loops,   $\varrho_{\rm 
 irr}(H)$,  shown   in   Fig.~\ref{fig1}.   The   increasing   annealing 
 temperature activates the hysterons with higher  coercivity.  Note  the 
 regions  with  negative  amplitudes  followed  by   ''regular''   ones. 
 Amplitudes  expressed   in   the   same,   arbitrary   units,   as   in 
 Fig.~\ref{fig3}.

\newpage
\pagestyle{empty}

\end{document}